\documentclass[12pt]{iopart}
\usepackage{iopams,texdraw}  
\begin{document}

\comment[H.~W. Diehl and M. Shpot: Lifshitz-point critical behaviour to
$O(\epsilon^2)$]{
   Lifshitz-point critical behaviour to
 ${\boldsymbol{O(\epsilon^2)}}$}

\author{H.~W. Diehl and M. Shpot}

\address{Fachbereich Physik, Universit{\"a}t
Essen, D-45117 Essen, Federal Republic of Germany}

\begin{abstract}
  We comment on a recent letter by L.~C. de~Albuquerque and M.~M.
  Leite (J.\ Phys. A: Math.\ Gen. {\bf 34} (2001) L327--L332), in which
  results to second order in $\epsilon=4-d+\frac{m}{2}$
  were presented for the critical exponents $\nu_{{\mathrm{L}}2}$,
  $\eta_{{\mathrm{L}}2}$ and $\gamma_{{\mathrm{L}}2}$ of
  $d$-dimensional systems at $m$-axial Lifshitz points. We point out
  that their results are at variance with ours. The discrepancy is due
  to their incorrect computation of momentum-space integrals. Their
  speculation that the field-theoretic renormalization group approach,
  if performed in position space, might give results different from
  when it is performed in momentum space is refuted.
\end{abstract}


\submitto{\JPA}

\section*{}
In a recent letter \cite{AL01}  de~Albuquerque
and Leite (AL) presented results to second order in
$\epsilon=4-d+\frac{m}{2}$ for the critical exponents
$\nu_{{\mathrm{L}}2}$, $\eta_{{\mathrm{L}}2}$ and
$\gamma_{{\mathrm{L}}2}$ of $d$-dimensional systems with $m$-axial
Lifshitz points.  For the special case $m=1$ of a uniaxial Lifshitz
point, these results were previously given in a (so far apparently
unpublished) preprint \cite{AL00}. The $\epsilon^2$ terms
AL found are at variance with ours \cite{DS00b,SD01}.

As an explanation for these discrepancies AL suggest the following.
Both we as well as AL employed a field-theoretic renormalization group
approach based on dimensional regularization. 
To compute the residua of the ultraviolet poles at $\epsilon
=0$, we found it convenient to perform (part of) the calculation in
position space. By contrast, AL worked entirely in momentum space.
They speculate \cite{AL01} ``that calculations performed in momentum
space and coordinate space are inequivalent, as far as the Lifshitz
critical behaviour is concerned''.

This speculation is untenable and a serious misconception, a fact
which should be obvious not only to readers with a background in field
theory. The reason simply is: at each step of the calculation one can
transform from momentum space to position
space and vice versa.

To become more specific, consider an $N$-point vertex function
$\Gamma^{(N)}({\boldsymbol{x}}_1,\ldots,{\boldsymbol{x}}_N)$ of a
dimensionally regularized, translationally invariant, renormalizable
Euclidean field theory, such as the $|\phi|^4$ theory for an $m$-axial
Lifshitz point considered both by us \cite{DS00b,SD01} and AL. In
momentum space the vertex functions have the form
$$
\tilde{\Gamma}^{(N)}({\boldsymbol{q}}_1,\ldots
,{\boldsymbol{q}}_N)\,(2\pi)^d\,
{\delta{\left(\sum_{i=1}^N{\boldsymbol{q}}_i\right)}}\;,
$$
where $\tilde{\Gamma}^{(N)}$ are conventional \emph{functions} of
$N-1$ independent momenta, e.g., ${\boldsymbol{q}}_1,\ldots,
{\boldsymbol{q}}_{N-1}$. The $\tilde{\Gamma}^{(N)}$ also depend on
$\epsilon$ (i.e., on $d$): they are \emph{meromorphic} in $\epsilon$, having
ultraviolet (uv) poles at $\epsilon=0$.  Since the issue is the $\epsilon$
expansion, these are the only poles we have to consider; possible
other uv poles at special values $\epsilon>0$ need not concern us here.
Likewise, we do not have to worry about possible infrared poles one
encounters in perturbation expansions about the Lifshitz point for a
fixed space dimension $d$, nor embark on a discussion of related
subtle questions like the appearance of perturbatively non-accessible
mass-shifts meromorphic in $\epsilon$, and on how these problems are avoided
in massive, fixed-$d$ renormalization schemes.

The Fourier back-transforms of the \emph{functions}
$\tilde{\Gamma}^{(N)}$ define \emph{generalized functions
  (distributions)}, which depend on the $N-1$ difference variables
${\boldsymbol{x}}_{j1}\equiv{\boldsymbol{x}}_j-{\boldsymbol{x}}_{1}$,
$1\le j \le N-1$. The same applies to each individual Feynman integral
contributing to $\Gamma^{(N)}$.

Field-theoretical renormalization group approaches such as the ones on
which AL's work and ours are based hinge on the possibility of
introducing a well-defined \emph{renormalized} theory by absorbing the
uv singularities of the theory's primitively divergent vertex
functions in a consistent manner through counter-terms that are
\emph{local in position space}.  In order that the renormalization
procedure can be interpreted as a re-parametrization, these
counter-terms must have the form of the (local) interactions appearing
in the original Hamiltonian, except for a finite number of admissible
additive ones.  Well-known mathematical renormalization theorems
\cite{Hep69,BM77b} ensure that this is the case, order by order in
perturbation theory.

Central to the proofs of such renormalization theorems is the
observation that the primitive uv singularities have a \emph{local}
structure in \emph{position space}. It is precisely this property that
is crucial for the renormalizability of the theory. It ensures that
the counter-terms, computed to a given order of perturbation theory,
provide the subtractions for all divergent subintegrations of the
Feynman graphs of the next higher order that are required to cancel
\emph{all those uv singularities that do not have the form of local
  counter-terms.}  Such nonlocal uv singularities occur indeed: for
instance, the graph
\raisebox{-8pt}{\begin{texdraw}
\drawdim pt \setunitscale 2.5   \linewd 0.3
\move(-7 -1.5)\rlvec(15 6)
\move(-7 1.5)\rlvec(15 -6)
\move(5 0)
\lellip rx:1 ry:3
\end{texdraw}} %
has momentum-dependent pole terms $\sim \epsilon^{-1}$ (involving
logarithms of momenta). These are due to the divergent subintegral
\raisebox{-2.2pt}{\begin{texdraw} \drawdim pt \setunitscale 2.5
    \linewd 0.3 \lellip rx:4 ry:1.8 \move(4 0)\rlvec(2 2) \move(4
    0)\rlvec(2 -2) \move(-4 0)\rlvec(-2 -2) \move(-4 0)\rlvec(-2 2)
\end{texdraw}}
; they do \emph{not} have the form of local counter-terms but cancel upon
making the appropriate subtraction for this subgraph.  (This
subtraction is produced by part of the one-loop counter-term $\propto
\phi^4$; see, e.g., Sec.\ 3.B of Ref.~\cite{Die86a}.)  Zimmermann's
forest formula \cite{Zim70} clarifies precisely which subtractions have
to be made for each individual Feynman graph.  The locality of the
counter-terms manifests itself in the fact that in the final
subtractions which must be made for superficially divergent graphs the
graph is \emph{shrunk to a point}.

What we have just explained has been known for decades and can be
found in standard textbooks on field theory. It is true that many
authors for computational reasons prefer the momentum representation
when explaining the renormalization procedure. Therefore the
significance of the uv singularities' local structure in position
space may escape the reader's attention if not properly emphasized.
However, a very clear exposition of the importance of this locality
is given already in one of the earliest, classic textbooks on
renormalization \cite{BS59}.

The renormalization procedure can be performed equally well in
momentum or position space. Utilizing dimensional regularization in
conjunction with minimal subtraction of poles is advantageous in that
one does not have to worry about how the regularization scheme and the
conditions for fixing the counter-terms translate upon Fourier
transformation: the scheme can be applied equally well in the momentum
or position representation. The upshot of these considerations is that
\emph{there is no way that AL's and our calculation can be both
  correct}.

The source of the discrepancies between AL's work and ours can be
traced back to the different results they find for the required
two-loop integrals. For example, our result for the integral
$I_3(p,k)$ defined in Eq.~(3) of AL's paper \cite{AL01} reads
\begin{equation}\label{eq:I3poles}
   I_3(p,k)=\frac{(2\pi)^{2d^*}}{\epsilon}\,{
\left[
\frac{j_\sigma(m)\,k^4}{16\,m\,(m+2)}
-\frac{j_\phi(m)\,p^2}{2\,(8-m)}
\right]}+O(\epsilon^0)\;,
\end{equation}
with
\begin{equation}\label{eq:jphi}
j_\phi(m)={\frac{{2^{10 + m}}\,{{\pi }^{6 + {\frac{3\,m}{4}}}}\,
     \Gamma({\frac{m}{2}})}{\Gamma( 2 - {\frac{m}{4}})\,
 {{\Gamma({\frac{m}{4}})}^2}}}{\int_0^\infty}{\mathrm{d}}
 \upsilon\,\upsilon^{m-1}\,
\Phi^3(\upsilon;m,d^*)\;,
\end{equation}
where
\begin{equation}\label{eq:Phi}
  \Phi(\upsilon;m,d^*)=
\int\frac{{\mathrm{d}}^{d^*-m}p}{(2\pi)^{d^*-m}}
\int\frac{{\mathrm{d}}^mk}{(2\pi)^m}\,
\frac{e^{i\,({\boldsymbol{p}}\cdot{\boldsymbol{e}}+
{\boldsymbol{k}}\cdot{\boldsymbol{\upsilon}})}}{p^2+k^4}
\end{equation}
is the scaling function associated with the free critical propagator
in position space (cf.\ Eq.~(13) of Ref.~\cite{DS00b}), at the upper
critical dimension $d^*=4+{m\over 2}$. Here ${\boldsymbol{e}}$ is a
unit $d^*-m$ vector, while ${\boldsymbol{\upsilon}}$ is an arbitrarily
directed $m$-vector.  The integral $j_\sigma(m)$ is similar to
$j_\phi(m)$, except that its integrand has an additional factor
$\upsilon^4$.

From AL's Eqs.~(11) and (18), we can infer their result for
$I_3(p,0)$; it reads
\begin{equation}\label{AL1}
I^{({\mathrm{AL}})}_3(p,0)=-\pi^{4+{m\over 2}}\,
\frac{\Gamma^2{\left({m\over 4}\right)}}{\Gamma^2{\left({m\over 2}\right)}}\,
\frac{1}{8-m}\,\frac{p^2}{\epsilon}+O(\epsilon^0)\;.
\end{equation}
To see that this cannot be correct, one must merely consider the
isotropic case $m=d=8-\epsilon$: for this, AL's result (\ref{AL1})
predicts a pole $\propto\epsilon^{-2}$, even though the pole part
$\propto p^2$ \emph{must vanish} because $\boldsymbol{p}$ has $d-m=0$
components. By contrast, our result (\ref{eq:I3poles})--(\ref{eq:Phi}) does not violate this
condition since $j_\phi(8)=0$. (See Sec.\ 4.5 and 4.4 of
Ref.~\cite{SD01} where we verified that our $\epsilon$-expansion
results for general values of $m$ reduce to known ones in both
isotropic case $m=d$ and $m=0$, respectively.)

AL realized the incorrectness of their findings for $m=8$. Yet they
seem to believe that the `approximations' they made in their
computation of $\ell\ge 2$ loop integrals do not lead to erroneous
results. Details of their approximations are described in Ref.~\cite{AL00}.
The crux of their method is `to impose the constraint'
${\boldsymbol{k}}_1=-2{\boldsymbol{k}}_2$
on the momenta of the internal integral
\begin{equation}\label{I2}                                      
  I_2({\boldsymbol{p}}_1+{\boldsymbol{p},\boldsymbol{k}}_1)=
\int\frac{{\mathrm{d}}^{d-m}p_2\,{\mathrm{d}}^{m}k_2}
{(p_2^2+k_2^4)[({\boldsymbol{p}}_1+
{\boldsymbol{p}}_2+{\boldsymbol{p}})^2+
({\boldsymbol{k}}_1+{\boldsymbol{k}}_2)^4]}
\end{equation}
of
\begin{equation}\label{I3}
  I_3(p,0)=\int\frac{{\mathrm{d}}^{d-m}p_1\,{\mathrm{d}}^{m}k_1}{p_1^2+k_1^4}
I_2({\boldsymbol{p}}_1+{\boldsymbol{p}},{\boldsymbol{k}}_1)\;.
\end{equation}
This amounts to modifying the momentum term
$({\boldsymbol{k}}_1+{\boldsymbol{k}}_2)^4$ of the last propagator in
Eq.~(\ref{I2}) to $k_2^4$. The error this introduces is given by the
analogue of the integral (\ref{I3}) one obtains through replacement of
$I_2({\boldsymbol{p}}_1+{\boldsymbol{p}},{\boldsymbol{k}}_1)$ by
the corresponding difference
$\delta I_2(.,k_1)\equiv I_2(.,{\boldsymbol{k}}_1)-I_2(.,0)$, namely
\begin{eqnarray}
  \delta I_2({\boldsymbol{p}}_1+{\boldsymbol{p}},k_1)&=&
\int\frac{{\mathrm{d}}^{d-m}p_2\,{\mathrm{d}}^{m}k_2}
{(p_2^2+k_2^4)[{\boldsymbol{p}}_1+
{\boldsymbol{p}}_2+{\boldsymbol{p}})^2+
({\boldsymbol{k}}_1+{\boldsymbol{k}}_2)^4]}\nonumber\\&&\qquad\times
\frac{k_1^4+4\,(k_1^2+k_2^2)\,{\boldsymbol{k}}_1\cdot{\boldsymbol{k}}_2
+6\,k_1^2\,k_2^2}{[({\boldsymbol{p}}_1+
{\boldsymbol{p}}_2+{\boldsymbol{p}})^2+k_2^4]}\;.
\end{eqnarray}
Now the pole term $\propto p^2/\epsilon$ of $I_3$ we are concerned
with corresponds to a logarithmic uv divergence $\sim p^2\,\ln\Lambda$
at the upper critical dimension ($\Lambda=$ cut-off). In order for AL's
approximation to be correct, $\delta I_2$ must have no contributions
that vary as $p^2\,p_1^{-2}$ or $p^2\,k_1^{-4}$ as $p_1\sim
k_1^2\sim\Lambda\to \infty$. As can be seen for instance by power
counting, this condition is \emph{not satisfied}. (Readers preferring
more mathematical scrutiny might want to compute
$\nabla_{\boldsymbol{p}}^2\,\delta I_2$ and study its behaviour for
large $p_1$ and $k_1$.) Accordingly, AL's approximation is unjustified
whenever $m\ne 0$. The same kind of approximations are employed by AL
for other $\ell \ge 2$ loop integrals.

In closing, let us outline how the pole term $\propto {p^2/ \epsilon}$
of $I_3$ given in Eq.~(\ref{I3}) can be recovered via a momentum-space
calculation. Using a Schwinger representation for each one of the
three propagators in Eqs.~(\ref{I2}) and (\ref{I3}), and performing
the Gaussian integrations over ${\boldsymbol{p}}_1$ and
${\boldsymbol{p}}_2$, we obtain
\begin{eqnarray}
  \label{eq:I3pole}
  I_3(p,0)&=&\pi^{d-m}{\int_0^\infty}\!{\mathrm{d}}x
{\int_0^\infty}\!{\mathrm{d}}y{\int_0^\infty}\!{\mathrm{d}}z\,
(xy+yz+zx)^{-\frac{d-m}{2}}\,
\nonumber\\&&\times
{\int}{\mathrm{d}}^mk_1{\int}{\mathrm{d}^m}k_2\,
{\mathrm{e}}^{
-\frac{xyz\,p^2}{xy+yz+zx}
-x\,k_1^4-y\,k_2^4-z\,|{\boldsymbol{k}}_1+{\boldsymbol{k}}_2|^4
}\;.
\end{eqnarray}
Next, we make the variable transformations $X=x/z$, $Y=y/z$ and
${\boldsymbol{K}}_{1,2}=z^{1/4}{\boldsymbol{k}}_{1,2}$, and take the
derivative $-\left.\partial/\partial p^2\right|_{p=1}$ inside the
integrals. The integration over $z$ can now be performed; it produces
the factor
$\Gamma(\epsilon)\,(1+X^{-1}+Y^{-1})^\epsilon
={1/ \epsilon}+O(\epsilon^0)$. Upon transforming to the variables
$s=1/X$ and $t=1/Y$, one finds that
\begin{eqnarray}
  \label{eq:Res}
{-\partial I_3(p,0)\over \partial p^2}
&=&
{\pi^{d^*{-}m}\over\epsilon}\,{\int_0^\infty}\!{\mathrm{d}\/}s
{\int_0^\infty}\!{\mathrm{d}\/}t\,(st)^{-{m\over 4}}\,(1+s+t)^{{m\over
    4}-3}
\nonumber\\&&\times
{\int}{\mathrm{d}}^mK_1{\int}{\mathrm{d}}^mK_2\;
{\mathrm{e}}^{-\frac{K_1^4}{s}-\frac{K_2^4}{t}-|K_1+K_2|^4}
+O(\epsilon^0)\;.
\end{eqnarray}
This is in conformity with Eqs.~(\ref{eq:I3poles})--(\ref{eq:Phi}).
To see this, note that the integral $j_\phi$ is
proportional to $\int{\mathrm{d}\/}^m\upsilon\,\Phi^3$. In momentum
space, this is a convolution of the form
$\int_{{\boldsymbol{k}}_1,{\boldsymbol{k}}_2}
\tilde{\Phi}_{{\boldsymbol{k}}_1}\tilde{\Phi}_{{\boldsymbol{k}}_2}
\tilde{\Phi}_{{\boldsymbol{k}}_1+{\boldsymbol{k}}_2}$. The
Fourier transform $\tilde{\Phi}_{{\boldsymbol{k}}}$ can be read off
from Eq.~(14) of Ref.~\cite{DS00b}; it involves a modified Bessel
function $K_\nu(k^2)$, for which we use the representation
\begin{equation}
  \label{eq:Bessel}
  k^{2\nu}\,K_\nu(k^2)=2^{\nu-1}{\int_0^\infty}\!
{\mathrm{d}\/}x\,x^{\nu-1}\,{\mathrm{e}\/}^{-x-\frac{k^4}{4x}}\;,
\end{equation}
with integration variables $x$, $y$ and $z$. Employing the transformations
$s=x/z$, $t=y/z$ and
${\boldsymbol{K}}_{1,2}=z^{-1/4}{\boldsymbol{k}}_{1,2}$, we perform
the integration over $z$. The result is the residuum of the pole
(\ref{eq:Res}).

To summarize: AL's results are incorrect  because of their unacceptable
approximations made
in computing Feynman diagrams. Their speculation that the field-theoretic
RG approach might yield different results depending on whether it is
performed in position or momentum space does not hold.

We gratefully acknowledge the support by the Deutsche
Forschungsgemeinschaft via the Leibniz programme Di 378/2-1.

\end{document}